\documentclass[aps,twocolumn,nofootinbib]{revtex4}
\usepackage[latin1]{inputenc}
\usepackage{epsfig}
\usepackage{graphicx}
\newcommand{\beq}{\begin{equation}}
\newcommand{\eeq}{\end{equation}}

\begin{document}

\title{Effect of immunization  through vaccination on
the SIS epidemic spreading model}

\author{Tânia Tomé and Mário J. de Oliveira}
\affiliation{Universidade de São Paulo, Instituto de Física,
Rua do Matão, 1371, 05508-090 São Paulo, SP, Brazil}

\begin{abstract}

We analyze the susceptible-infected-susceptible model
for epidemic spreading in which a fraction of the
individuals become immune by vaccination. This process
is understood as a dilution by vaccination, which
decreases the fraction of the susceptible individuals.
For a nonzero fraction of vaccinated individuals, the model 
predicts a new state in which the disease spreads
but eventually becomes extinct.
The new state emerges when the fraction of vaccinated
individuals is greater than a critical value. The model
predicts that this critical value increases as one
increases the infection rate reaching an asymptotic value,
which is strictly less than the unity. Above this 
asymptotic value, the extinction
occurs no matter how large the infection rate is.

\end{abstract}

\maketitle

\section{Introduction}

In the second edition of his book on the prevention
of malaria \cite{ross1911},
Ross developed an analytical theory of dynamics
of infectious diseases through the use of
differential equations of the first order in time
\cite{ross1911,ross1915}.
He called the approach the theory of happenings,
understood as events that transform
the condition of an individual such as a disease or a
vaccination. 
The time evolution of the number $z$ 
of infected individuals in a population $p$ that
remains constant, he represented by the equation
\beq
\frac{dz}{dt} = cz(p-z) - rz,
\label{1}
\eeq
and called $c$ the infection rate and $r$ a factor that is
related to the reversion or recovery of the individuals.
He also solved this equation and found an `S' shaped
curve for $z$ and a bell shaped curve of new cases,
given by $cz(p-z)$.

A simple epidemic model involving only the infection 
process was later developed by Bailey
\cite{bailey1950,bailey1957}.
In this model all susceptible individuals eventually become
infected. The evolution equation
for the number $y$ of susceptible individuals was written as
\beq
\frac{dy}{dt} = - \beta y(n+1-y),
\label{2}
\eeq
where $\beta$ is the infection rate and $n$ is the number of
susceptible individuals at $t=0$. He also developed
a stochastic version of this model 
where the number of susceptible and infected individuals
are treated as stochastic variables. 
This simple model was extended by Weiss and Dishon 
through the introduction of a recovery process
\cite{weiss1971}. Symbolically, the two processes
of their model were represented by
\beq
S \stackrel{\alpha\,\,}{\to} 2I, \qquad\qquad
I \stackrel{\beta}{\to} S,
\label{2a}
\eeq
where $\alpha$ is the infection rate and $\beta$ the recovery rate.
The evolution equation for the number $x$
of infected individuals was written as
\beq
\frac{dx}{dt} = \alpha x(N-x) -\beta x,
\label{2b}
\eeq
which is identical to the Ross equation (\ref{1}).
They also developed a stochastic approach to this model
where the numbers of susceptible and infected individuals
are treated as stochastic variables.

The model represented by the reaction equation (\ref{2a}) 
is known as the susceptible-infected-susceptible (SIS) model
and the equations (\ref{2b}) or (\ref{1}) describe the
deterministic version of this model
\cite{kryscio1989,mollison1995,keeling2008,tome2020a}.
The model represented
by the first reaction equation alone is known as the
susceptible-infected (SI) model and the equation (\ref{2})
describes the deterministic version of the model.
The stochastic versions are understood generically as a 
random walk on a space where the axes are the number of
the various classes of individuals \cite{nisbet1982,tome2020b}.

The models for epidemic spreading and in general models
that describe population dynamics can also be defined as
having a spatial structure and evolving through a stochastic
dynamics. When the spatial structure is represented by a
lattice, these models are referred to as stochastic lattice models
\cite{harris1974,grassberger1983,grassberger1983a,ohtsuki1986,
satulovsky1994,durrett1995,antal2001,dahmen2003,tome2009,souza2010,
tome2011,souza2013,tome2015,ruziska2017,ana2017}.
In these models each site of a lattice is in one of several
possible states. The change of the state of a given site follows
a stochastic dynamics in accordance with the  
processes represented by the reaction equations.
An interesting feature of the stochastic lattice models
is that the deterministic version as well as the stochastic
version referred to above may be obtained from 
the stochastic lattice models through an approximation
where the correlations are entirely or partially
neglected \cite{satulovsky1994,tome2009,tome2011,souza2013}.
This is the procedure we will follow here
to obtain the deterministic equations of the SIS 
epidemic model with vaccination.

A stochastic lattice model which is 
equivalent to the SIS model was in fact advanced by Harris,
who called it a {\it contact process} \cite{harris1974}.
Each site of the lattice
is occupied by a particle or is empty. A particle
is created on an empty site, with a certain creation rate,
if a nearest neighbor is occupied. Particles
disappear spontaneously with a certain annihilated rate.
By identifying a particle with an infected individual
and an empty site with a susceptible individual,
the contact process becomes equivalent to the 
stochastic lattice SIS model. 

Here we are concerned with the analytical description of
the process of vaccination on the SIS model and the effects 
it produces on the spreading of the infectious disease.
Vaccination is a powerful tool for the mass prevention of 
infection by inducing in the body an immune reaction
which in turn generates
immunity to infections \cite{keeling2013}.
The immunity may be only partial but 
here we assume that the immunity is permanent.
To properly 
describe this process we add a third class of individuals
to represent the vaccinated individuals.
In addition to the class of susceptible (S) and infected
(I), we consider also the class of vaccinated (V) 
individuals, which have acquired permanent immunization
through vaccination. We consider also that
the process of vaccination starts at the beginning
of the spreading of the disease and that in the long
run a fraction $k$ of the individuals are vaccinated.

The SIS model presents two behaviors according to
the strength of the infection rate. For small 
infection rate there is no spreading. When the 
infection rate is greater that a certain
value, the spread of the epidemic sets in, and 
the disease becomes endemic, or persistent, 
which means that in the long run there is a nonzero fraction of
infected individuals. In accordance with the model
of vaccination analyzed here, if the
fraction of vaccinated individuals is large enough
the epidemic eventually disappears, or that the
disease becomes extinct, meaning that the
fraction of infected vanishes in the long run. 
The effect of vaccination is to turn the `S' 
shaped curve of infected into a bell shaped curve.
Our analysis shows that the critical value
above which the disease disappears increases with
the infection rate, reaching an asymptotic value
which we found to be strictly less that the unity.
In other words, if the fraction of vaccinated is
larger than this asymptotic critical value,
the extinction of the disease
occurs no matter how large the infection rate is.

The process of vaccination is usually approached by a
spontaneous process represented by the reaction equation
S$\to$V which occurs with a certain vaccination rate 
\cite{kribs2000,alexander2006,shaw2010,wang2016,
pires2017,khanjanianpak2020,kuga2021}.
If we apply this approach to a deterministic
model, the whole population will eventually
become infected. We wish here to describe 
a process of vaccination in which only a
fraction $k$ of the individuals will be vaccinated.
The way in which we introduce the vaccination process is
explained in the following.

\section{Vaccination process}

The epidemic model with vaccination that we analyze is defined
on a lattice, each site of which is occupied by an
individual that can be in various conditions with respect
to the infectious disease. We first analyze the case in 
which the individuals are sedentary, remaining forever in
th same sites of the lattice. Here it suffices to consider
two classes of individuals. The class of individuals that
have been vaccinated (V) and the class of individuals that
have not been vaccinated (N). The vaccination process 
that we wish to set up is such that in the long
run a certain fraction $k$ of individuals are vaccinated. 
A natural way of setting up this process is to choose
randomly a certain set ${\cal A}$ of $k$ sites of the lattice
whose individuals will be vaccinated.
Initially, all sites are in state N.

The sites of ${\cal A}$ are subject to the process
N$\to$V which we assume to occur with a rate $a$.
The sites of the complementary set of sites
will never be in state V, remaining forever in state N.
As the sites of ${\cal A}$ are randomly chosen a priori
and do not change in time, the set ${\cal A}$ is understood
as comprising a {\it quenched disordered} set. 
In accordance with the reaction N$\to$V,
the probability $P_V$ that a site of ${\cal A}$
is in state V evolves in time according to
\beq
\frac{d}{dt}P_V = a P_N,
\eeq  
where $P_N$ is the probability that the site is in state N.
As $P_N = 1 - P_V$, this equation becomes an equation in $P_V$
whose solution is
\beq
P_V = 1-e^{-at},
\eeq
where we are considering that at $t=0$ the site is
in state N.
From this result it follows that the fraction $v$ of 
all sites of the lattice that are in the state V
varies in time according to 
\beq
v = k(1-e^{-at}),
\label{4}
\eeq
where $k$ is the number of sites of ${\cal A}$ and
is the fraction of vaccinated individuals in the long run.

We now suppose that the individuals are not sedentary but move
on the lattice without keeping their positions.
As a consequence, the sites belonging to ${\cal A}$ change with time.
If we focus on a specific
site which in a certain instant of time is in state V,
we see that now this site may change to the state N,
a impossibility in the case of a quenched disorder set. 
Therefore, if we consider a {\it specific site} of the
lattice, which is the case when we set up the stochastic
process on a lattice, we have to consider also the process
V$\to$N in addition to the process N$\to$V. Denoting by
$\gamma$ and $\beta$ the transition rates of the former
and latter processes, the time evolution of $P_V$ is
\beq
\frac{d}{dt}P_V = - \gamma P_V + \beta P_N.
\label{6}
\eeq
Again, as $P_N=1-P_V$, this equation can be solved with
the solution
\beq
P_V = \frac{\beta}{\beta+\gamma}[1 - e^{-(\alpha+\beta) t}].
\label{5}
\eeq
We next assume that $P_V$ is the same for all sites and
as a consequence it is equal to the fraction $v$ of 
vaccinated sites. We also assume that $v$ has the same
form of the previous quenched approach given by (\ref{4}).
Comparing (\ref{5}) with (\ref{4}), we see that
the rates $\alpha$ and $\beta$ are related to the
vaccination rate $a$ and to the fraction $k$ of
vaccinated individuals by $\beta = ak$ and $\gamma = a(1-k)$.
Using these relations, equation (\ref{6}) is written as
\beq
\frac{d}{dt}P_V = a(k-P_V),
\label{8b}
\eeq
and (\ref{5}) as
\beq
P_V = k(1-e^{-at}).
\label{8a}
\eeq

The first approach to vaccination, which is associated
to the quenched disorder, is appropriate to describe a
situation in which the individuals are fixed in space,
an example being the spread of an epidemic occurring
in an orchard where the trees are attached to the ground.
In the epidemic spreading occurring in a community
of moving individuals, the appropriate approach
is that associated with the process of
{\it dilution by vaccination} described by the equation
(\ref{8b}). Here we consider just the second approach,
applying it to the SIS model. It is worth mentioning that
the stationary properties of the SIS model with the
first approach to vaccination are identified with
contact process under quenched dilution
\cite{moreira1996,vojta2005,oliveira2008,wada2017}.

\section{Model}

We present now the SIS model under vaccination appropriate
to describe an epidemic spreading on a community of moving
individuals. The sites of a lattice are occupied by three
types of individuals: susceptible (S), infected (I), or
vaccinated (V). A susceptible becomes infected through an
auto-catalytic process S$\to$I with rate $b$. An infected
become susceptible through a spontaneous reaction I$\to$S
with rate $c$. These two processes comprise the ordinary
SIS model. In accordance with the dilution approach to
vaccination explained above, the processes involving the
vaccinated individuals are as follows. An S or an I become
V with rate $ak$ and V becomes S with rate $a(1-k)$.
These processes are represented by the reaction equations
S$\to$ V and I$\to$V, each one occurring with rate 
$\beta=ak$, and V$\to$S occurring with rate $\gamma=a(1-k)$,
where $a$ is the rate of vaccination and $k$ is the fraction
of individuals to be vaccinated.

These considerations allow us to set up a stochastic process
on a lattice, which is described by a master equation that
governs the time evolution of the probability of the
global state of the system \cite{tome2015L}. From the
master equation we may derive the evolution equations
for the various marginal probabilities \cite{tome2015L}.
The evolution equations
for the probabilities $P_S$, $P_I$, and $P_V$ of a site
being occupied by a susceptible, an infected, and a vaccinated
individuals, respectively, are
\beq
\frac{d}{dt} P_S = - b P_{SI} + c P_I - \beta P_S + \gamma P_V,
\label{6b}
\eeq
\beq
\frac{d}{dt} P_I =   b P_{SI} - c P_I - \beta P_I,
\label{6c}
\eeq
\beq
\frac{d}{dt}P_V = - \gamma P_V + \beta (P_S+P_I),
\label{6a}
\eeq
where $P_{SI}$ is the probability of a site being in state S
and one of its neighbors in state I.
These equations are not all independent because $P_S+P_I+P_V=1$.

Replacing $P_S+P_I=1-P_V$ in equation (\ref{6a}), it becomes
\beq
\frac{d}{dt}P_V = a(k- P_V),
\label{8e}
\eeq
which is in accordance with equation (\ref{8b}).
The solution of
this equation with the initial condition with no vaccinate
individuals is
\beq
P_V = k(1-e^{-at}),
\label{8f}
\eeq
which is in accordance with (\ref{8a}).
Equations (\ref{8e}) and (\ref{8f}) are a consequence of
the independence of each site with respect to the
dynamics of the vaccinated individuals, a relevant
property of the present model. In
general, the probability that in a set of $n$ sites,
$m$ sites are in state V and the remainder are not
in state V is the product $P_V^{\,m}(1-P_V)^{n-m}$.

Similarly we may write the evolution equations for the
pair correlations $P_{SI}$, which is given by
\[
\frac{dP_{SI}}{dt} = - c P_{SI} - b (1-r)P_{SI} -b r P_{ISI}
+ cP_{II} + b r P_{SSI}
\]
\beq
- 2\beta P_{SI} + \gamma P_{IV},
\label{30a}
\eeq
where $r=(\kappa-1)/\kappa$ and
$\kappa$ is the coordination number of the lattice,

The set of evolutions equations are to be solved for the initial condition
such that all sites are in the S state except a small number of
them which is in the I state. The set of equations are not closed
equations and to solve them we resort in two approximations.

\begin{figure*}
\centering
\epsfig{file=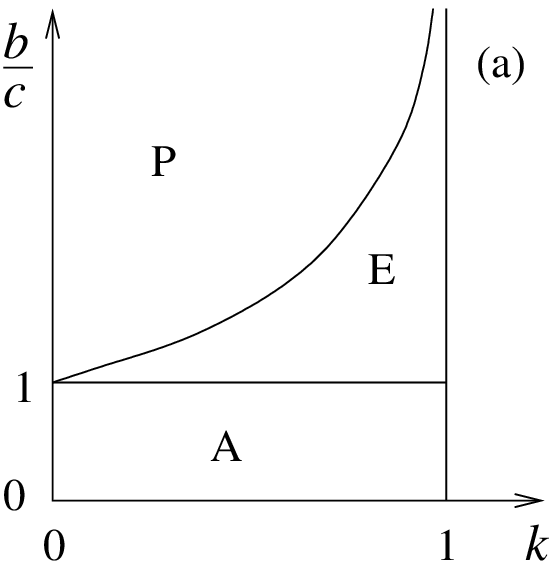,width=4.5cm}
\hskip3cm
\epsfig{file=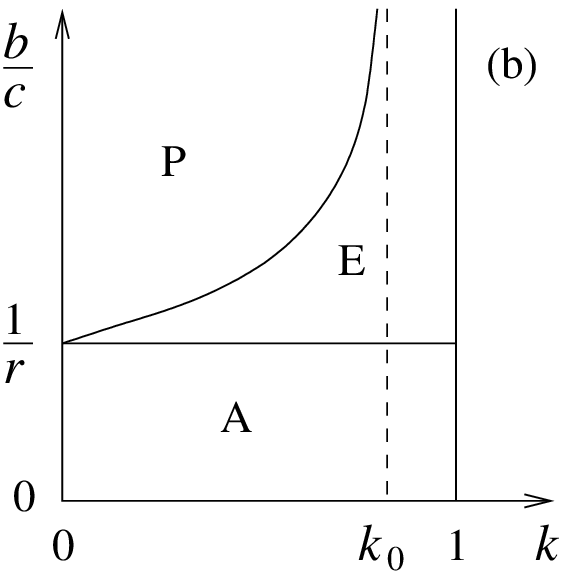,width=4.5cm}
\caption{Diagram showing the behavior of the SIS model
with vaccination, according to the first approximation
(a) and to the second approximation (b) where $b$ is
the infection rate, $c$ is the recovery rate, and $k$
is the fraction of individuals that have acquired
permanent immunization by vaccination, where
$k_0$ is strictly less than 1. In the region A there is
no spreading of the disease, whereas in P and E the
disease spreads. In P it becomes persistent, with a
nonzero fraction of infected, and in E it becomes
extinct, with a zero fraction of infected individuals.
The figures were drawn for a very small value of the
vaccination rate $a$.}
\label{sisv}
\end{figure*}

\section{First approximation}

We consider here an approximation in which the probability $P_{SI}$ is
replaced by the product $P_S P_I$, which is understood as a full
break of the correlations. Using the notation
$x=P_S$, $y=P_I$, and $v=P_V$, the equation (\ref{6c}) becomes
\beq
\frac{dy}{dt} =   b x y- (c+\beta) y,
\label{7}
\eeq
where $x=1-y-v$ and $v$ is considered to be a function of
time given by
\beq
v=k(1-e^{-at}),
\label{8c}
\eeq
where $a=\beta+\gamma$. 
This equation is to be solved
for an initial condition such that $y$ is small at $t=0$. 

One solution of this equation corresponds to the absence of
infected individuals, that is, $y=0$. To determined
the stability of this solution, we linearize (\ref{7})
around this solution to get
\beq
\frac{dy}{dt} =   b (1-v) y- (c+\beta) y.
\label{7b}
\eeq
At the initial times the fraction of vaccinated individuals
is zero, $v=0$, and equation (\ref{7b}) becomes
\beq
\frac{dy}{dt} =   (b - c-\beta) y.
\eeq
The solution of this equation tell us that
$y$ decreases if $b<b_1$, where
\beq
b_1 = c + a k.
\eeq
Therefore, if $b<b_1$ there is no spreading of the epidemic,
the epidemic is absent, a state we denote by A.

In the opposite case, $b>b_1$, $y$ grows initially,
and in the long run it reaches an asymptotic value.
We envisage two behaviors for the time evolution of
$y$. In one of them, $y$ stops increasing and 
asymptotically reaches a nonzero value, indicating
that the disease becomes endemic, or persistent,
a state we call P. In the other, $y$ stops increasing,
then decreases and reaches a zero value which means
that the disease becomes extinct, a state we call E.

The nonzero asymptotic value of $y$ is given by the
stationary solution of equation (\ref{7}). For large times,
we replace $v$ by its asymptotic value $v=k$
and the equation (\ref{7}) becomes
\beq
\frac{dy}{dt} =   b (1-k-y) y- (c+\beta) y.
\label{7a}
\eeq
In the stationary state
\beq
y = (1-k) - \frac{c+\beta}{b}.
\eeq
This solution occurs as long $y>0$, that is, as long as
$b\geq b_2$, where
\beq
b_2=\frac{c + a k}{1-k}.
\eeq

Alternatively, we may consider the stability of the
solution $y=0$ in the regime of large times in which
case we replace $v$ in the linearized equation (\ref{7b})
by $k$, and equation (\ref{7b}) becomes 
\beq
\frac{dy}{dt} =   b (1-k) y- (c+\beta) y.
\eeq
From this equation we conclude that the solution $y=0$
is stable if $b<b_2$.

In figure \ref{sisv}a we show the phase diagram
in the variables $b$ and $k$
which displays the various states 
A, E and P according to the present approximation.
The line $b=b_1$ determines the AE transition
line and $b=b_2$ determines the EP transition
line. According to this approximation,
the effect of immunization through vaccination
is to cause the disappearance of the disease,
or its extinction, when the fraction $k$
of vaccinated individuals is large enough.

\begin{figure*}
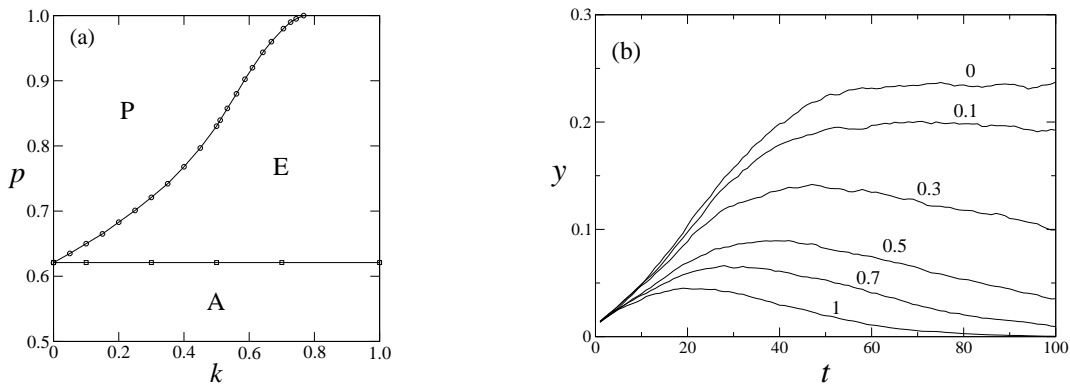

\centering
\epsfig{file=pXk.eps,height=5cm}
\hskip2cm
\epsfig{file=tk.eps,height=5cm}
\caption{(a) Phase diagrams obtained by numerical simulations
in the plane $p$ versus $k$. The states are persistent (P),
extinct (E) and no spreading (A).
(b) Fraction $y$ of infected individuals as a function
of time $t$ obtained by numerical simulations for $p=0.7$
and several values of the fraction $k$ of vaccinated
individuals. For $k=$0.3, 0.5, 0.7 and 1, $y$ vanishes
in the long run.}
\label{simul}
\end{figure*}

\section{Second approximation}

Now we wish to consider a second approximation 
in which the correlations are partially broken
\cite{tome2011}.
The approximation we use now is to break the three-site 
correlations by using the replacements:
$P_{ISI}=P_{SI}^2/P_S$ and $P_{SSI}=P_{SS}P_{SI}/P_S$.
The equation (\ref{30a}) becomes
\[
\frac{dP_{SI}}{dt} = - c P_{SI} -b (1-r) P_{SI}
- b r\frac{P_{SI}^{\,2}}{P_S}
+ cP_{II} + br \frac{P_{SI}P_{SS}}{P_S}
\]
\beq
- 2\beta P_{SI} + \gamma P_{IV},
\eeq

We use the notations $P_S=x$, $P_I=y$, $P_V=v$, $P_{SI}=u$,
the approzximations $P_{IV}=yv$ and $P_{SV}=xv$, 
and the identities $P_{II}+u+yv=y$ and $P_{SS}+xv+u=x$
to write the equation above in the form
\[
\frac{du}{dt} = - c u -b (1-r) u - b r\frac{u^2}{x} + c(y-u-yv) 
\]
\beq
+ br \frac{u}{x} (x-u-xv) - 2\beta u + \gamma yv,
\label{9}
\eeq
Equation (\ref{9}) together with equation (\ref{6c}),
which we write in the form
\beq
\frac{dy}{dt} = - c y + b u -\beta y,
\label{10}
\eeq
constitute a close set of equations for $y$ and $u$,
and we recall that $v$ is a function of time given by
(\ref{8c}), and that $x=1-y-v$.

The trivial solution of these equations is $y=0$, $u=0$,
which corresponds to the absence of spreading. To determine
its stability, we linearize equations (\ref{9})
to find
\beq
\frac{du}{dt} = - c u -b (1-r) u + c(y-u-yv) 
+ br u(1-v) - 2\beta u + \gamma yv,
\label{13a}
\eeq
At the initial times $v=0$ and this equations become
\beq
\frac{du}{dt} = - c u -b (1-2r) u + c(y-u) - 2\beta u,
\label{12a}
\eeq
The stability analysis using equations (\ref{12a}) and 
(\ref{10}) leads to conclude that the state without the
spreading of disease is stable, or that $y$ will not grow
if  $b<b_1$ where
\beq
b_1 = \frac{2(c+ak)^2}{2rc+(2r-1)ka}.
\eeq
That is, the epidemic will not spread if $b<b_1$.

For long times, $v$ approaches $k$ and the linearized
equation (\ref{13a}) becomes
\beq
\frac{du}{dt} = - c u -b (1-r) u + c(y-u-yk) 
+ br u(1-k) - 2\beta u + \gamma yk.
\label{15a}
\eeq
The stability analysis using equations (\ref{15a})
and (\ref{10}) leads to conclude that
the solution corresponding to extinction state
is stable if $b<b_2$ where
\beq
b_2 = \frac{2(c+ak)}{2r-k-rk}.
\eeq
When $b>b_2$, there emerges a state where the disease becomes
endemic.

The phase diagram predicted by this second approximation
is qualitatively similar to that of the first approximation
as can be seen in figure \ref{sisv}b. However, there is an
important quantitative difference. If the fraction $k$ of
vaccinated individuals is larger that $k_0$, which is
strictly less that the unity, the disease disappears,
or becomes extinct no matter how large the rate of
infection is. From the line $b=b_2$ we find 
\beq
k_0 = \frac{2r}{1+r}.
\eeq
For a square lattice $r=3/4$ and $k_0=6/7=0.857$.

\section{Numerical simulations}

We have simulated the SIS model with vaccination
on a square lattice with periodic boundary conditions
with $L\times L$ sites, with $L$ ranging from 10 to 
80 and 10$^6$ Monte Carlo steps.
The transition rules that we used are as 
as follows. At each time step, a site of the lattice
is chosen at random. With probability $\varepsilon$,
the process of vaccination is carried out, otherwise,
with the complementary probability $1-\varepsilon$, the
SIS processes are preformed. In the first case,
if the chosen site is in state S or I, it becomes V with
probability $k$, and if it is in state V, it becomes S with
probability $1-k$. In the second case, if the chosen site
is in state I then it becomes S with probability $1-p$. If
it is in state S, then one of its nearest neighbors is
chosen at random, and if the chosen neighboring site is in state I,
then the chosen site becomes I with probability $p$.

We have initially determined the stationary values
of the fraction $y$ of infected individuals.
The vanishing of $y$ determines the transition 
line between the states E and P as shown in the
phase diagram of figure \ref{simul}a in the
variables $p$ and $k$ for $\varepsilon=0.01$.
We recall that $p$ is related to the rate of infection
and $k$ is the fraction of vaccinated individuals
in the long run. To find the transition line between
the A and E states, we have calculated $y$ as a function
of time with an initial condition with all sites
in state S except one which is in the state I.
The transition point occurs when $y$ begins to 
increase with time as one increases the value
of $p$ at a fixed value of $k$.

The line $k=0$ corresponds to th ordinary SIS model which
has a transition from the disease free state A to
the persistent state P occurring at $p=0.62246$
\cite{tome2015L}. From this point, two transition lines
emerge, as shown in figure \ref{simul}a.
The first line separates the state A and the
extinction state E, where the disease becomes extinct.
In the state E, the fraction of infected individuals
increases, reaches a maximum and then decreases and
vanishes in the long run as can be seen in figure
\ref{simul}b. A second line separates the E and P
states. This line ends at a point along the line
$p=1$ at the value of $k$ equal to $k_0=0.760(2)$ for
$\varepsilon=0.01$.
From these results we draws the following conclusions.
For a given value of $p$, and thus of the infection rate,
there is a critical value of the fraction $k$ of vaccinated
individuals above which the disease becomes extinct. 
The critical value increases with the infection rate
and reaches the value $k_0$ at the maximum possible
value of the infection rate which corresponds to $p=1$.
If the value
of $k$ is larger than $k_0$ the disease becomes extinct
no matter how large the infection rate is, a result that
we have found also within the second approximation.

We have also analyzed the critical behavior of the
fraction of infected individuals, which is understood
as the order parameter of the P state. We calculated
the critical exponent $\beta$ along the transition
line EP and found values of $\beta$ which
are consistent with that of that predicted by contact process,
namely $\beta=0.583$ \cite{tome2015L}.
These results indicate that the whole EP line
belongs in the universality class of the contact process,
also known as the direct percolation universality class.

\section{Conclusion}

We have analyzed an extension of the SIS model 
by the inclusion of the process of vaccination. To this
end a third class of individuals, the ones that
have acquired permanent immunization by vaccination,
were added to the
classes of susceptible and infected. 
The process of vaccination is understood
as a dilution of the system by transforming the
susceptible into vaccinated, and thus decreasing the 
fractions of the susceptible and infected. For a
nonzero fraction of vaccinated individuals,the model 
predicts a new state in which the disease spreads
but eventually becomes extinct, in addition to
the two states of the ordinary SIS model, the
absence of epidemic spreading and the endemic state, 

As one increases the fraction $k$ of vaccinated individuals, 
there is a critical value of $k$ above which the
disease becomes extinct. The critical value increase
with the strength of the infection rate. However, as
the infection rate increases without bounds, the
critical fraction approaches a value $k_0$ which is
strictly less than the unity. This result says that
no matter how large the infection rate is,
the disease disappears if the fraction of 
individuals is larger than $k_0$.
From the numerical simulations of the model on a
square lattice,
we have found $k_0=0.760(2)$ for $\varepsilon=0.01$,

A relevant feature of the process of immunization 
by vaccination that we used here is that the dynamics
of the vaccination is independent of the dynamics
of the other processes. In addition the dynamics related
to a specific site is independent of other sites.
Let us denote by N a site that is either in state S or I
and ask for the probability that each site of the lattice
is in a certain given state, either V or N. This probability
is equal to the product $P_V^{\,n}P_N^{\,m}$, where
$P_N=1-P_V$, $n$ is the number of sites in state V,
and $m$ is the number of sites in state N.

\section*{Acknowledgement}

We wish to take this opportunity to congratulate Robert Ziff on his
seventieth birthday.


\end{document}